\documentclass[twocolumn,amsmath,amssymb,floatfix,prb,superscriptaddress,showpacs]{revtex4-1}
\usepackage{amsmath,amssymb,natbib,bm,graphicx,url,epsfig,psfrag}
\usepackage{color}
\usepackage{ulem}

\newcommand{\red}[1]{{\color{red}{#1}}}

\begin{document}
\title{Topological dephasing in the $\nu=2/3$ fractional Quantum Hall regime}
\author{Jinhong Park}
\thanks{Present affiliation: Department of Condensed Matter Physics, Weizmann
Institute of Science, Rehovot 76100, Israel}
\affiliation{Department of Physics, Korea Advanced Institute of
Science and Technology, Daejeon 305-701, Korea}
\author{Yuval Gefen}
\affiliation{Department of Condensed Matter Physics, Weizmann
Institute of Science, Rehovot 76100, Israel}
\author{H.-S. Sim}  \email{hssim@kaist.ac.kr}
\affiliation{Department of Physics, Korea Advanced Institute of
Science and Technology, Daejeon 305-701, Korea}
\date{\today}

\begin{abstract}
We study dephasing in electron transport through a large quantum dot (a Fabry-Perot interferometer) in the fractional quantum Hall regime with filling factor $2/3$. In the regime of sequential tunneling, dephasing occurs due to electron fractionalization into counterpropagating charge and neutral edge modes on the dot. 
In particular, 
when the charge mode moves much faster than the neutral mode,
and at temperatures higher than the level spacing of the dot, 
electron fractionalization combined with
the fractional statistics of the charge mode leads to the dephasing selectively suppressing $h/e$ Aharonov-Bohm oscillations but not $h/(2e)$ oscillations, resulting in oscillation-period halving.  
\end{abstract}



\maketitle

\section{Introduction}
A fractional quantum Hall (QH) system of filling fraction $\nu$ has edge channels that support fractional charges obeying  fractional braiding statistics~\cite{Wen_review}. At $\nu = 2/3$, the edge states are decomposed into a $\nu_c^{\textrm{edge}}=2/3$ charge mode and a counterpropagating neutral mode~\cite{Kane_PRL,Kane_PRB}.
They originate from renormalization of two counterpropagating charge modes~\cite{Neutral.MacDonald, Neutral.Johnson}, $\nu_1^{\textrm{edge}}=1$ and $\nu_2^{\textrm{edge}}=-1/3$, and stabilize at low temperature under strong disorder. 
Neutral modes have attracted much attention, as they are charge neutral and carry energy. 
They have been recently detected through shot noise measurements~\cite{Neutral.Bid}, and
their properties such as energy and decay length have been extensively 
studied~\cite{Overbosch,Neutral.Rosenow,Neutral.Takei,Neutral.Viola,Gross,Gurman,Venkatachalam,Neutral.Altimiras,Inoue2,Neutral.Wang,Meier,Kamenev, Bishara, Bonderson}.

Electron interaction is a dominant source of dephasing at low temperature~\cite{Hansen}. 
It leads to electron fractionalization~\cite{Pham,Jagla} in quantum wires; an electron, injected into a wire, splits into constituents (spin-charge separation, charge fractionalization), showing reduction of interference visibility or dephasing~\cite{Neutral.LeHur}.
Interestingly, when the wire is finite, the constituents recombine after bouncing at wire ends, resulting in coherence revival~\cite{Neutral.Kim}.
Fractionalization was detected~\cite{Neutral.Steinberg} in a non-chiral wire, and studied in the integer QH edge~\cite{Levkivskyi,Neutral.Berg,Neutral.Horsdal,Neutral.Neder,Neutral.Bocquillon,Neutral.Inoue1}.


Coherent transport, as well as dephasing, can be tackled through the study of low energy dynamics at the edge. This is particularly important in the context of   the fractional QH regime. The
 present study implies that 
 the presence of neutral modes could be a dominant source of 
 dephasing. Note that neutral modes have been observed
 in almost all fractional QH systems~\cite{Inoue2}. At the same time there is no uncontested observation of anyonic interference oscillations in the pure Aharonov-Bohm regime of a fractional QH interferometer\red{~\cite{Ofek}}.


The present study of the $\nu = 2/3$ QH regime emphasizes two dephasing mechanisms by fractionalization of an electron into charge and neutral components, {\it plasmonic dephasing} and {\it topological dephasing}.
Concerning  the plasmonic
 dephasing mechanism, the overlap between the
 plasmonic parts of the charge and neutral components
 decreases with time, as the two components propagate
 with different velocities in the opposite directions.
 The resulting  dephasing is similar to the  plasmonic
 dephasing that takes place in a quantum wire or in  integer
 QH edges.
On the other hand, the topological dephasing is a new mechanism unnoticed so far. It occurs because the zero-mode parts of the components, satisfying fractional statistics, may braid with thermally excited anyons.
Thermal average of the resulting braiding phase 
leads to dephasing that occurs only in the interfering processes characterized by particular values of topological winding numbers.



\begin{figure}[b]
\includegraphics[width=\columnwidth]{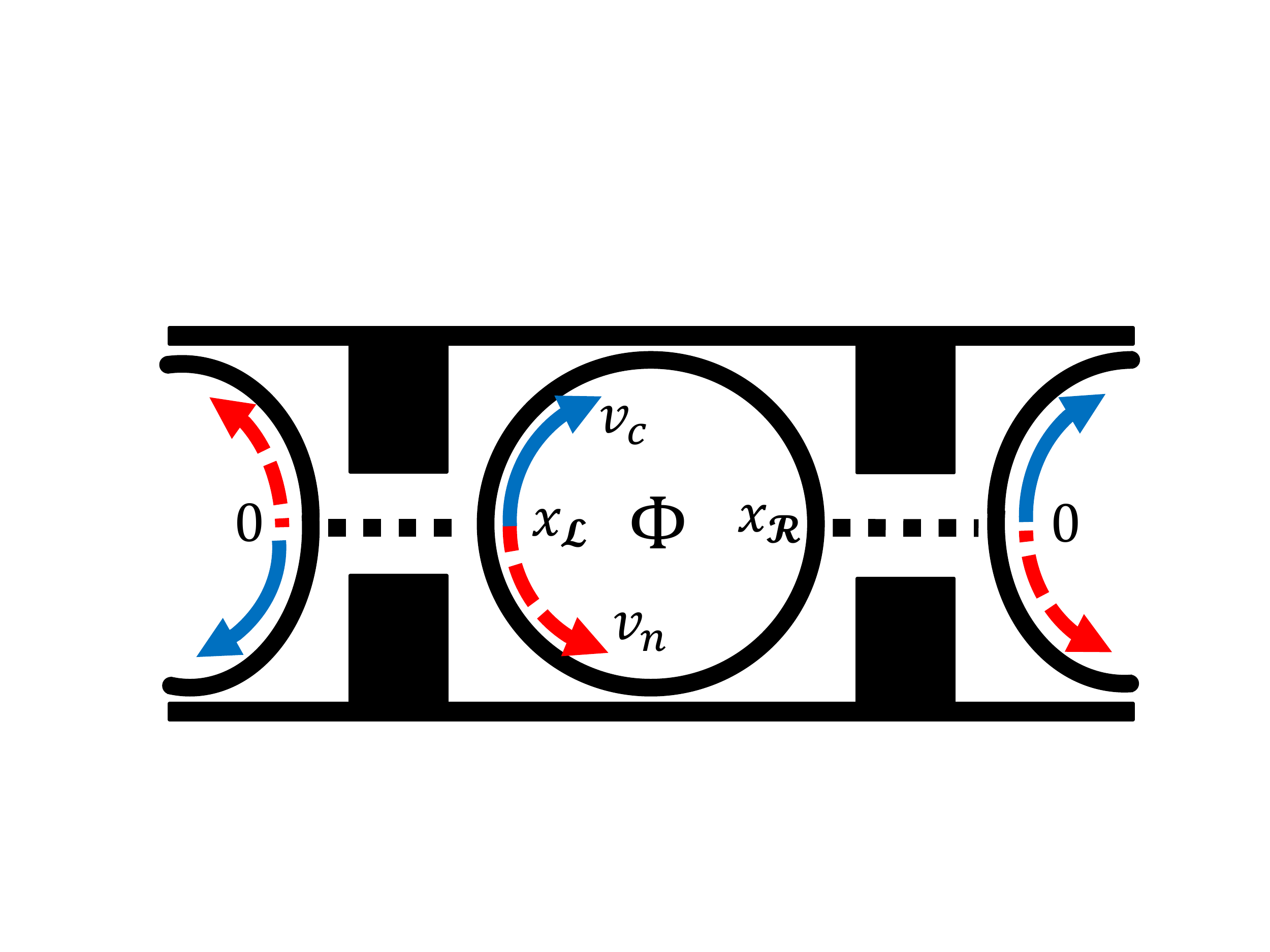}
\caption{(Color online) A large quantum dot (Fabry-Perot interferometer) in the fractional QH regime of $\nu = 2/3$, coupled to lead edge states of $\nu = 2/3$ (black solid lines) through quantum point contacts (QPCs) at $x_\mathcal{L,R}$. Electron (rather than  fractional quasiparticle) tunneling occurs through the QPCs (dotted lines). Following the tunneling, each electron (and the hole left behind in the lead edge) fractionalizes into a charge component propagating at velocity $v_c$ (solid blue arrow) and a neutral component counterpropagating at velocity $v_n$ (dashed red). 
The magnetic flux in the dot area is $\Phi$. 
}\label{Setup1}
\end{figure}

Our analysis addresses the AB oscillation of differential conductance $\mathcal{G}$ through a quantum dot (QD) in the $\nu = 2/3$ QH regime. We focus on linear response of electron sequential tunneling into the QD. $\mathcal{G}$
is decomposed into the harmonics of the AB flux $\Phi$ in the QD,
\begin{eqnarray}
\mathcal{G} = \frac{e^2}{h} \sum_{\delta p=0,1,2,\cdots} \mathcal{G}_{\delta p} \cos ( 2 \pi \delta p \frac{\Phi}{\Phi_0}), \label{g_decomp}
\end{eqnarray} 
where $\Phi_0 \equiv \hbar c / |e|$ is a flux quantum; see Fig.~\ref{Setup1}.
Semiclassically, $\delta p$ represents the relative winding number of a fractionalized charge component, around the circumference $L$ of the QD, between two interfering paths:
an electron, after tunneling into the QD, fractionalizes into charge and neutral components; see Fig.~\ref{Setup1}. The charge (neutral) component has propagation velocity $v_{c(n)}$, spatial width $L_{T,c (n)} \equiv  \hbar v_{c (n)} / (2 \pi k_B T \delta_{c (n)})$ at temperature $T$, level spacing $E_{c (n)} \equiv 2\pi \hbar v_{c (n)} / L$,
and scaling dimension $\delta_c = 3/4$ ($\delta_n = 1/4$) in the electron tunneling operator at low temperatures.
$\mathcal{G}_{\delta p}$ is determined by the overlaps of the components of the same kind between two interfering paths of relative charge winding $\delta p$.

We find two mechanisms suppressing $\mathcal{G}_{\delta p \ne 0}$, the plasmonic dephasing and the topological dephasing; the former (latter) involves plasmon (zero-mode) parts of the components.  
In the plasmonic dephasing, 
$\mathcal{G}_{\delta p}$ is contributed from the two interfering paths whose charge components overlap maximally between the paths. But, their neutral
components 
overlap only partially between the interfering paths, reducing $\mathcal{G}_{\delta p}$;
similar dephasing occurs in other fractionalizations~\cite{Neutral.LeHur,Neutral.Kim}.
The topological dephasing additionally occurs, but depending
on $\delta p$, in  contrast to the other known mechanisms.
When $v_{c} \gg v_{n}$~\cite{Wan,Lee}, 
the first harmonics $\mathcal{G}_{\delta p=1}$ is suppressed at 
$k_B T > E_n / (4\pi^2 \delta_n)$ (namely, $L > L_{T,n}$).
It is because the charge component gains thermally fluctuating fractional braiding phase of $\pi N_c$ (leading to $e^{i \pi N_c} =\pm 1$), 
while it winds once ($\delta p = 1$) around $N_c$ electronic or anyonic thermal excitations on the QD edge or in the bulk.
By contrast, the second harmonics $\mathcal{G}_{\delta p=2}$ is not affected by the topological dephasing (as braiding phase $\pi N_c \delta p$ and $(\pm 1)^{\delta p}=1$ are trivial) and dominates $\mathcal{G}$, resulting in $h/(2e)$ AB oscillations.
These above findings occur in both the regimes of strong disorder and weak disorder in the edge of the QD. Note that the topological dephasing does not occur in the Coulomb dominated regime~\cite{Ofek, Rosenow} where Coulomb interactions between the bulk and edge of the QD is strong, as discussed later.




\section{Setup and Hamiltonian}
The $\nu=2/3$ QD is coupled to two lead edges via quantum point contacts (QPCs)~\cite{Neutral.Furusaki}; see Fig.~\ref{Setup1}. The Hamiltonian is $H=H_\textrm{D} + H_\mathcal{L} + H_\mathcal{R} + H_\textrm{T}$. $H_\textrm{D}$ describes the edge of the QD, while $H_\mathcal{L(R)}$ the $\nu = 2/3$ left (right) lead edge. Each edge consists of the bosonic mode $\phi_1$ ($\nu_1^{\textrm{edge}}=1$) and the counterpropagating $\phi_2$ ($\nu_2^{\textrm{edge}}=-1/3$). $\phi_{i=1,2}$ supports charge $e \nu_i^{\textrm{edge}}$ and satisfies $[\phi_i(x), \phi_{i'}(x')] = i\pi \nu_i^{\textrm{edge}} \text{sgn}(x-x')\delta_{ii'}$ at positions $x, x'$.
Introducing the charge mode $\phi_c \equiv \sqrt{3/2}(\phi_1+\phi_2)$ (supporting charge $2e/3$) and the neutral mode $\phi_n \equiv (\phi_1 + 3\phi_2)/\sqrt{2}$,  one writes~\cite{Kane_PRL,Kane_PRB}
\begin{eqnarray} \label{Hamiltoniandot}
H_\textrm{D} &=& \frac{\hbar}{4\pi} \int^L_0 dx [v_c (\partial_x \phi_c)^2 + v_n(\partial_x \phi_n)^2+v\partial_x\phi_c\partial_x \phi_n] \nonumber \\
&+&\int^L_0 dx [\xi (x) \exp(i\sqrt{2}\phi_n) + \text{H.c.}].
\end{eqnarray}
Disorder-induced tunneling amplitude $\xi(x)$ between $\phi_1$ and $\phi_2$ is modeled by a Gaussian random variable with mean zero and variance $\overline{\xi^{*}(x)\xi(x)}=W\delta (x-x')$. For a finite range of bare parameters, $\phi_c$ and $\phi_n$ decouple~\cite{Kane_PRL} at low temperatures, rendering $v$ irrelevant. $H_\mathcal{L,R}$ is written similarly to $H_{\textrm{D}}$, except $\int^L_0 \to \int^\infty_{-\infty}$ in Eq.~\eqref{Hamiltoniandot}.
Note that we ignore the Coulomb interaction between the bulk and edge of the QD, considering that the QD size is large enough~\cite{Rosenow}. 


The QPCs are almost closed, so electron  tunneling is facilitated.
Renormalization group analysis~\cite{Kane_PRL,Kane_PRB} indicates four equally most relevant electron tunneling operators between the electron field operators, $\Psi_{\pm} (x_\alpha) = e^{i\sqrt{3/2}\phi_c (x_\alpha)}e^{\pm i \phi_n (x_\alpha)/ \sqrt{2}} / \sqrt{2\pi a}$ at $x_{\alpha = \mathcal{L,R}}$ on the QD, and
$\Psi_{\alpha, \pm}(0)$ on lead edge $\alpha$; $a$ is an ultra-violet cutoff and $\Psi_{\alpha, \pm}$ has the same form as $\Psi_{\pm}$. So the tunneling Hamiltonian is
$H_\textrm{T} = \sum_{\alpha = \mathcal{L},\mathcal{R}} \sum_{i,j=\pm} [t_{\alpha ij} \Psi_{\alpha,i}^{\dagger} (0) \Psi_{j} (x_\alpha) + \text{H.c.}]$, where $t_{\alpha ij}$ is the tunneling strength.

\section{Topological dephasing}
We show that at $\nu = 2/3$, fractionalization and fractional statistics cause the topological dephasing. 
We address the 
number operator $N_{c (n)}$ of charge (neutral) mode at the QD edge,
\begin{eqnarray} \label{OldnumberNewnumber}
\frac{1}{3}N_c = N_1 - \frac{1}{3} N_2, \quad \quad N_n = N_2-N_1,
\end{eqnarray}
defined through the zero-mode parts of $\phi_{1,2}$ (see Appendix~\ref{AppenNumber}).
The number operator $N_{1(2)}$ of $\phi_{1(2)}$ is an integer since $e$ and $-e/3$ are the elementary charges of $\phi_{1,2}$; $N_c$ is an integer measuring charge 
excitations in the units of $e/3$ ($N_c = 1$ for a quasiparticle of charge $e/3$; $N_c = 3$ for an electron).

A quasiparticle of charge $e/3$ at position $x$ on the QD edge is written as $e^{i \phi_c(x) / \sqrt{6}}e^{\pm i \phi_n(x) / \sqrt{2}}$~\cite{Neutral.Viola}.
Consider clockwise exchange of two such quasiparticles.
Since $[\phi_{c}(x), \phi_{c}(x')] = i\pi  \text{sgn}(x-x')$,
the exchange of the two charge components results in statistical phase $\pi/6$,
\begin{equation}
e^{\frac{i}{\sqrt{6}} \phi_c(x)} e^{\frac{i}{\sqrt{6}} \phi_c(x')} = e^{\pm i \frac{\pi}{6} \textrm{sgn}(x'-x)} e^{\frac{i }{\sqrt{6}}\phi_c(x')} e^{\frac{i }{\sqrt{6}}\phi_c(x)}.
\label{Braid}
\end{equation}
So, after the charge component of the electron operator $\Psi_\pm$ winds once clockwise around $N_c$ charge-mode excitations on the edge, a phase 
$3\times N_c \times 2 \times \pi/6 = \pi N_c$ is gained~\cite{Chamon1}. Here, 3 means the number of charge components forming $\Psi_\pm (x)$, and $2$ refers to braiding (double exchanges).
Similarly, the exchange of the neutral components of the two quasiparticles 
leads to exchange phase $-\pi/2$. So the neutral component of $\Psi_\pm (x)$ gains $\pm 1 \times N_n \times 2\times (-\pi/2)= \mp \pi N_n$, after winding once around $N_n$ neutral-mode excitations; the number of the neutral components of $\Psi_\pm$ is $\pm 1$.

\begin{figure}
\includegraphics[width=.9\columnwidth]{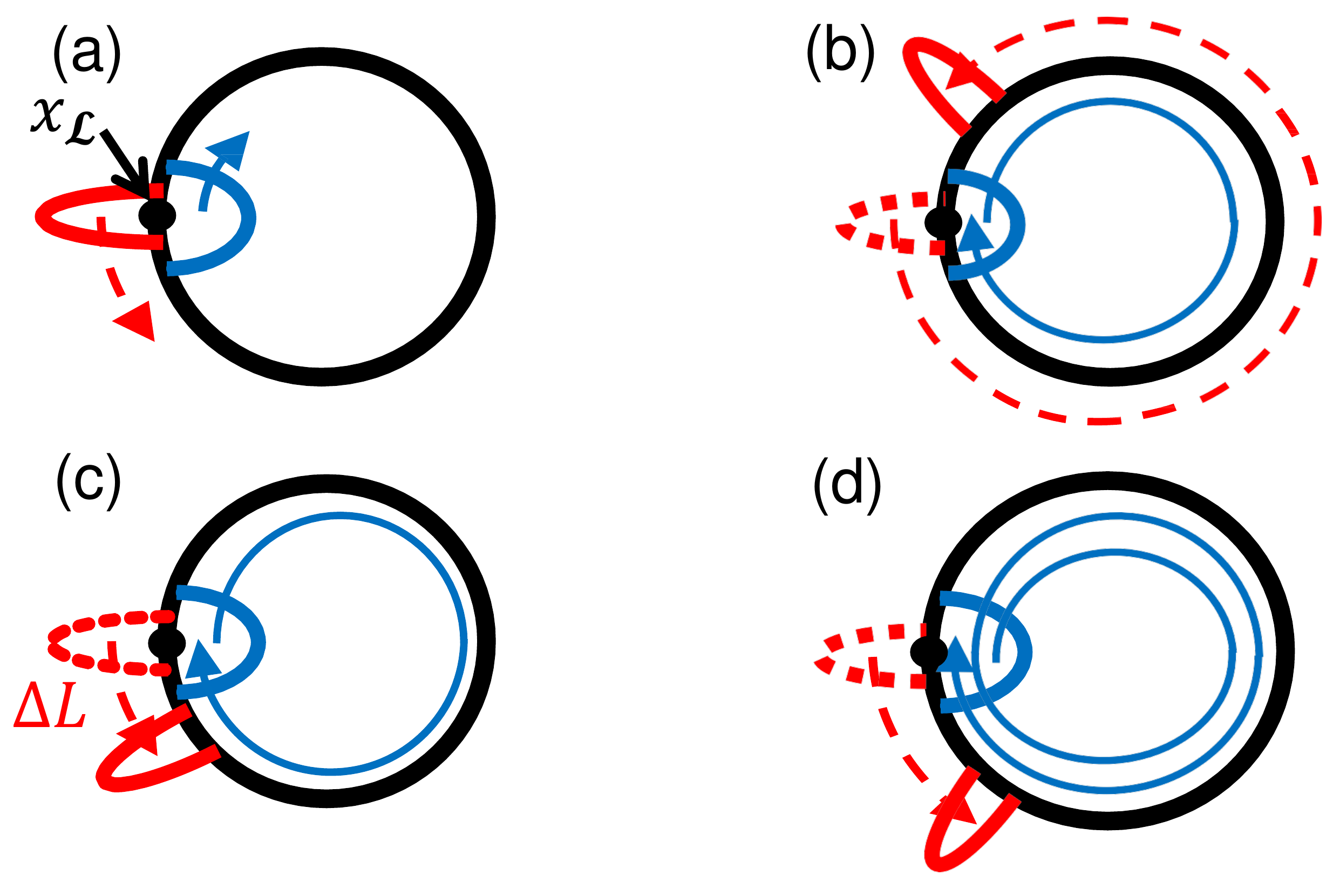}
\caption{(Color online) Dynamical processes involving different winding numbers
of the charge ($p$) and neutral ($q$) components. 
(a) $(p,q) = (0,0)$: an electron injected at $x_\mathcal{L}$ fractionalizes into
charge (moving along blue solid arrows) and neutral (red dashed) components.
(b) $(p,q) = (1,0)$. For $v_c \gtrsim v_n$, the charge (neutral) component
arrives at $x_\mathcal{L}$, after winding once around the QD, $p=1$ (almost once, $q=-1$). The interference of relative winding numbers $(\delta p,\delta q) = (1,-1)$ between (a) and (b) contributes to $\mathcal{G}_{\delta p =1}$. Reduced overlap between the neutral components of (a) and (b) leads to plasmonic dephasing.
For $v_c \gg v_n$, the dynamics is depicted for (c) $(p,q)=(1,0)$ and  
(d) $(p,q)=(2,0)$. 
The charge component winds once in (c) and twice in (d), while the neutral component moves little by $\Delta L$ in (c) and $2\Delta L$ in (d)  
(hence  mainly $q = 0$). 
The interference between (a) and (c) 
suffers from
topological dephasing with odd $\delta p + \delta q$. The interference between (a) and (d) is immune to topological dephasing. 
}\label{Interferencepaths2}
\end{figure}

This has implications on the dynamics of an electron
which enters into the QD and then fractionalizes. When $v_c \gg v_n$, there is a process where the charge component of the electron winds once around the QD, while the neutral component moves very little.
In terms of the winding numbers of the charge and neutral components, $p$ and $q$,
this process is denoted by $(p,q)=(1,0)$. 
This process interferes with that of no winding $(p',q') = (0,0)$, contributing to the $h/e$ harmonics $\mathcal{G}_{\delta p =1}$; see Fig.~\ref{Interferencepaths2}. 
The relative winding numbers between the two interfering paths are $(\delta p = p-p'=1, \delta q = q - q'=0)$, and the net braiding phase gained from that winding around $N_c$ charge and $N_n$ neutral excitations on the edge is $\pi (N_c \delta p  \mp  N_n \delta q) = \pi N_c$.
Since $N_c$ is an integer, thermal fluctuations of quasiparticle (or electron) excitations on the edge 
give rise to fluctuations of the braiding phase factor $e^{i \pi N_c} = \pm 1$ [$+$ ($-$) for even (odd) $N_c$], suppressing the $h/e$ harmonics.
This topological braiding-induced dephasing also occurs due to thermal quasiparticle or electron fluctuations in the bulk; see Appendix~\ref{Appendix_Quasiflucbulk} for quasiparticle fluctuations in the bulk.
Note that this topological dephasing is utterly different from a dephasing mechanism at zero temperature, arising when quasiparticles travelling along an edge change internal degrees of freedom within the bulk (e.g., Ref.~\cite{Stern}).



By contrast, the main contribution to the $h/(2e)$ harmonics $\mathcal{G}_{\delta p=2}$
comes from $(\delta p, \delta q)$ = (2,0). In this case, the braiding phase factor $e^{\pi i (N_c \delta p \mp N_n \delta q)} = 1$, regardless of $N_c$ being even or odd. Hence, $\mathcal{G}_{\delta p=2}$ is immune to
the topological dephasing. In general, such dephasing occurs only with odd $\delta p + \delta q$, since the fluctuating $N_c \pm N_n$ is always even; see Eq.~\eqref{OldnumberNewnumber}.

When $v_c \simeq v_n$, the topological dephasing does not occur, since $\delta p = -\delta q$ and $N_c \pm N_n$ is even.


The above arguments hold for the pure AB regime
 (or for the intermediate regime between pure AB and
 Coulomb-dominated regimes). An apt question is to what extent this analysis holds for the Coulomb dominated regime. When an electron of a given energy
 enters  the QD by the process of  sequential tunneling, it occupies
 a certain orbital state of the QD edge, satisfying energy
 conservation.  In the pure AB regime, the area enclosed by the orbit (hence, the AB
 phase assigned to  the orbit) is not modified when the number of
 quasiparticles or electrons in the bulk of the QD
 fluctuates thermally:  edge-bulk interactions are negligible. Hence,
 such thermal   
 fluctuations affect only the braiding phase
 gained by the electron, leading to topological dephasing.
 By contrast, in the  Coulomb-dominated regime, the
 fluctuations are fully screened by the edge, reflecting the effect of   edge-bulk
 interaction.  This  screening
 leads to modification of the area  of the orbit, hence it modifies  the
 AB phase of the orbit. This change of the AB phase exactly
 cancels out the change of the braiding phase caused
 by the thermal fluctuations. It follows that  topological
 dephasing disappears in the Coulomb-dominated regime.

\section{Sequential tunneling}
We compute $\mathcal{G}$ in Eq.~\eqref{g_decomp} to the order 
of sequential tunneling,
\begin{equation} 
\mathcal{G} \simeq \frac{e^2}{\hbar} c_g \tilde{\gamma} k_B T  \sum_{j=\pm, \alpha} \int_{-\infty}^{0} dt F (t)\text{Im} \, G_j(x_\alpha, x_\alpha; t), \label{conductance11}
\end{equation}
where $\tilde{\gamma} = \gamma_{\mathcal{L}}\gamma_{\mathcal{R}}/( \gamma_{\mathcal{L}}+\gamma_{\mathcal{R}})$, $\gamma_\alpha \propto |t_{\alpha ij}|^2$ is the (renormalized) electron tunneling rate between the QD and lead edge $\alpha = \mathcal{L,R}$, 
$G_j(x_{\alpha}, x_{\alpha}; t) \equiv \big\langle \big [\Psi_{j}^{\dagger}(x_{\alpha},t), \Psi_{j}(x_{\alpha},0) \big ]
\big\rangle$ is the Green function describing the time ($t$) evolution of the fractionalized  components of an injected electron described by $\Psi_{j}(x_\alpha,t=0)$, and $c_g$ is a constant.
The start and end positions of the Green function $G_j(x_{\alpha}, x_{\alpha}; t)$ coincide, since the Green function describes the sequential tunneling.
The injection leaves a hole behind on the lead edge.
$F(t) = (\pi k_B T t/ \hbar) \sinh^{-2 (\delta_c + \delta_n)} (\pi k_B T t / \hbar)$
accounts for the fractionalization of the hole.
For the detailed derivation of Eq.~\eqref{conductance11}, see Appendix~\ref{secAppendix_seq}.


$\mathcal{G}_{\delta p}$ comes from the interference between two processes of relative charge winding number $\delta p$.
At $k_B T > E_{c}/(2\pi^2)$, the charge component has spatial width $L_{T,c}< L$. Then, $G_j(x_\alpha, x_\alpha; t)$ contributes to $\mathcal{G}_{\delta p}$ mainly around the times $\delta p L / v_{c}$, at which the charge component arrives at the initial injection point $x_\alpha$ after winding $\delta p$ times around $L$; the neutral component winding times $ \delta q L / v_{n}$ are much less important, 
because of the scaling dimensions $\delta_c=3 \delta_n > \delta_n$.
We focus on $\mathcal{G}_{\delta p = 1}$ and $\mathcal{G}_{\delta p = 2}$, as
they involve the shorter times of $\delta p L / v_c$, are more robust against the dephasing discussed below, hence are much larger than $\mathcal{G}_{\delta p \ge 3}$  at $k_B T \gg E_{c}$.
At $k_B T \gg E_{c}$,
we compute $\mathcal{G}_{\delta p = 1}$ and $\mathcal{G}_{\delta p = 2}$ analytically in the absence of disorder and interaction ($W=0$, $v=0$), and also in the strong disorder regime, based on a finite-size bosonization~\cite{Haldane1,Loss,Geller,Eggert,Delft,Neutral.Kim} and a semiclassical approximation (see Appendix~\ref{appendix_semi}).

\section{Clean regime}
We first deal with the regime of $W=0$ and $v=0$ [see Eq.~\eqref{Hamiltoniandot}] and
then discuss the regime of weak disorder and weak intermode interaction.
We  treat various contributions to dephasing quantitatively for the two cases of $v_c \gtrsim v_n$ and $v_c \gg v_n$.

When $v_c \gtrsim v_n$ and $k_B T \gg E_c$, only the  plasmonic dephasing is important.
The dominant contribution to $\mathcal{G}$ comes from the $h/e$ harmonics.
With the additional condition of $k_B T \gg  \hbar v_n / (L -\Delta L)$, we obtain 
\begin{eqnarray} \label{conductancevcsimvn}
\mathcal{G}_{\delta p = 1} \propto
\tilde{\gamma} L  (k_BT)^3 \exp(-\frac{L}{L_{T,c}}-\frac{\Delta L}{L_{T,n}}-\frac{L - \Delta L}{L_{T,n}}), 
\end{eqnarray}
where $\Delta L =L v_n/v_c$.
We explain two processes, whose interference dominates $\mathcal{G}_{\delta p = 1}$.
In one process [Fig.~\ref{Interferencepaths2}(a)], an electron tunnels from lead edge  $\mathcal{L}$ into the QD and fractionalizes at $x_{\mathcal{L}}$ at time $t_{1}'=0$, while at $t_2' = - L/v_c$ in the other [Fig.~\ref{Interferencepaths2}(b)]. The charge components of the two processes interfere at $x_{\mathcal{L}}$ at $t = 0$, contributing to $\mathcal{G}_{\delta p = 1}$, after respective windings $p = 0$ and $1$.
At that time, the distance between the neutral components is $L - \Delta L$, leading to partial overlap, hence, to the third factor $\exp(-(L - \Delta L)/{L_{T,n}})$ of $\mathcal{G}_{\delta p = 1}$.
The tunneling leaves a hole behind on $\mathcal{L}$,
which also fractionalizes into charge and neutral components (not shown in Fig.~\ref{Interferencepaths2}). The partial overlap at $t = 0$ between the two charge components from the holes created at $t_1'$ and $t_{2}'$, and that between the two neutral components, lead to the first two exponential factors of $\mathcal{G}_{\delta p = 1}$ in Eq.~\eqref{conductancevcsimvn}, respectively. 
In the other limit of $v_c \gg v_n$, both  plasmonic and  topological dephasings are crucial. 
There are two interfering processes for  $\mathcal{G}_{\delta p = 1}$ shown in  Figs.~\ref{Interferencepaths2}(a) and  \ref{Interferencepaths2}(c),
and for $\mathcal{G}_{\delta p = 2}$ in Figs.~\ref{Interferencepaths2}(a) and \ref{Interferencepaths2}(d).
When $k_B T \gg E_c$, we obtain 
\begin{eqnarray}
\mathcal{G}_{\delta p = 1} &\propto&  
\tilde{\gamma} L  (k_BT)^3 \exp{(-\frac{L}{ L_{T,c}}-\frac{\Delta L}{L_{T,n}}-\frac{\Delta L}{L_{T,n}} 
+\frac{ (\Delta L)^2}{L L_{T,n}})}   \nonumber\\  && \times \exp[-\frac{( L - \Delta L)^2}{ L L_{T,n}}]  \label{conductancevcggvn1}   \\
\mathcal{G}_{\delta p = 2} 
&\propto& 
\tilde{\gamma} L  (k_BT)^3 \exp[-2(\frac{L}{L_{T,c}}+\frac{\Delta L}{L_{T,n}}+\frac{\Delta L}{L_{T,n}})]. \label{conductancevcggvn2}
\end{eqnarray}
The first three exponential factors of $\mathcal{G}_{\delta p = 1}$ and $\mathcal{G}_{\delta p = 2}$ result from   plasmonic dephasing, as those of Eq.~\eqref{conductancevcsimvn}.
The third factor has a form different from that of Eq.~\eqref{conductancevcsimvn} of $v_c \gtrsim v_n$, as
the interfering neutral components in the QD are now $\Delta L$ apart in space.
The factor $2$ in the arguments of $\mathcal{G}_{\delta p = 2}$ arises from the double winding. Another exponential factor  $\exp[ ( \Delta L)^2/ (L L_{T,n})]$ of $\mathcal{G}_{\delta p = 1}$ comes from the plasmonic part of the neutral component; it is cancelled out with zero-mode contributions in Eqs.~\eqref{conductancevcsimvn} and \eqref{conductancevcggvn2} and also in other cases~\cite{Neutral.Kim}.

The last suppression factor in Eq.~\eqref{conductancevcggvn1}, $\exp[-(L- \Delta L)^2/ (L L_{T,n})]$, represents  the  topological dephasing, arising from the zero-mode parts of $G_{j}(x_{\alpha}, x_{\alpha};t)$.
The process in  Fig.~\ref{Interferencepaths2}(c) (where the center of the neutral component hardly moves, while the charge component winds once around $L$) interferes with that of Fig.~\ref{Interferencepaths2}(a), contributing to  $(\delta p, \delta q) = (1,0)$.
As discussed around Eq.~\eqref{Braid}, this interference with $\delta p + \delta q = 1$ is suppressed by the thermally fluctuating braiding phase factor of $e^{i \pi (\delta p N_c + \delta q N_n)} = \pm 1$. The suppression factor is interestingly determined by the spatial tail (or  finite $L_{T,n}$) of the zero-mode part of the neutral component. 
The tail indicates
that the neutral component can quantum mechanically wind once more than the semiclassical number $q$ of the center; the quantum mechanical winding is well defined by the Poisson  formula  (see Appendix~\ref{appendix_top}). 
Hence, from the processes in Figs.~\ref{Interferencepaths2}(a) and \ref{Interferencepaths2}(c), interference with the total relative winding number of $\delta p + (\delta q + 1)$ can occur. As $\delta p + \delta q + 1$ is even, this interference avoids the topological dephasing, dominantly contributing to $\mathcal{G}_{\delta p=1}$, but it is reduced by the separation $L - \Delta L$ of the neutral components of $\delta q + 1$ relative windings.
This explains the factor $\exp[-(L- \Delta L)^2/ (L L_{T,n})]$; the exponent is quadratic in $L- \Delta L$, since it originates from the zero-mode part~\cite{Neutral.Kim}.

We point out that the topological dephasing occurs when $L > L_{T,n} $ ($k_B T > E_n / (4 \pi ^2 \delta_n)$), as seen in the last exponential factor in Eq.~\eqref{conductancevcggvn1}.
In contrast, the plasmonic dephasing occurs when $k_B T \gg E_c$.
Note that we choose the condition of $k_B T \gg E_c$ for the derivation of  
Eq.~\eqref{conductancevcggvn1}, to show both of  the plasmonic dephasing and the topological dephasing simultaneously.

\begin{figure}[tpb]
\includegraphics[width=\columnwidth]{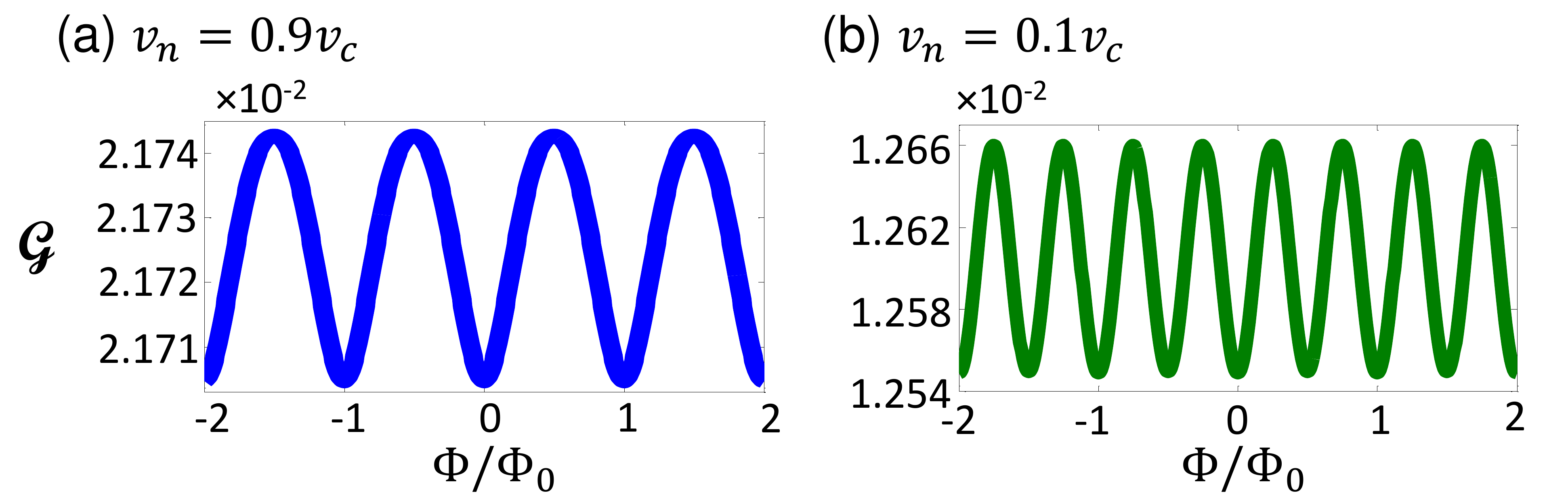}
\caption{(Color Online) 
Topological dephasing and period halving. Shown are 
Aharonov-Bohm oscillations of $\mathcal{G}$ for (a) $v_{n}  = 0.9v_{c}$ (period $\Phi_0$) and for (b) $v_{n} = 0.1 v_{c}$ (period $\Phi_0/2$) at $k_B T=E_c / 20$ (blue curve). $\mathcal{G}$ is measured in units of $e^2 \tilde{\gamma} a / (h^2 v_c^{3/4} v_n ^{1/4})$ and $L=200a$. 
}\label{conductance3} 
\end{figure}

Because of  the topological dephasing, $\mathcal{G}_{\delta p = 2}$ is much larger than $\mathcal{G}_{\delta p = 1}$ when $v_c \gg v_n$; $\exp (- (L - \Delta L)^2 /L L_{T,n})$ is much smaller than the other factors.
As a result, $\mathcal{G}$ shows $h/(2e)$ AB oscillations. In Fig.~\ref{conductance3}, we numerically compute $\mathcal{G}$ for both $v_c \gtrsim v_n$ and $v_c \gg v_n$ without employing the semiclassical approximation.
The result for $v_c \gg v_n$ demonstrates the topological dephasing and consequent period halving even at $k_B T < E_c$.

So far, we have discussed the regime of no intermode interaction ($v= 0$) and no disorder ($W=0$). The argument of the regime holds also in the regime of weak intermode interaction and weak disorder, with slight modifications. 
In this regime, the plasmonic part of the neutral component decays, together with the diffusive spreading of the plasmonic part of the charge component~\cite{Kane_PRB}.
These slightly modify the plasmonic dephasing (the first three 
dephasing factors in Eqs.~\eqref{conductancevcsimvn}$-$\eqref{conductancevcggvn2}, but do not affect  the topological dephasing.
Note that the weak disorder regime is realized when the renormalization of $W$ stops by temperature $T$ or QD size $L$ before going to the strong disorder regime, and a weak intermode interaction
 occurs in a dot, when the Coulomb interaction between the charge modes is larger than the confining potential (see Appendix~\ref{appen_Coulomb} and 
Refs.~\cite{Neutral.Wang, Chamon2}).
In recent experiments~\cite{Gurman}, neutral modes are measured with QDs of size $4 \, \mu$m$^2$, implying that the intermode interaction is sufficiently weak in the QDs.

\section{Strong disorder regime}
We show that   Eqs.~\eqref{conductancevcsimvn}$-$\eqref{conductancevcggvn2} hold in the strong disorder regime of a QD edge without any modification. 
In this regime, the neutral component is totally decoupled with the charge component ($v=0$ in Eq.~\eqref{Hamiltoniandot})~\cite{Kane_PRL}. 


We start with the diagonalized form of $H_\textrm{D}$ 
$H_\textrm{D} = \int_{0}^{L} dx [v_c (\partial_x \phi_c)^2 / (4\pi) + v_n  \tilde{\psi}^{\dagger}i\partial_x \tilde{\psi}]$.
This form is obtained from Eq.~\eqref{Hamiltoniandot}, where the effect of disorders is included.
Here, $\tilde{\psi}(x)  \equiv (e^{i(\tilde{\chi}+\tilde{\phi}_{n})/\sqrt{2}}, e^{i(\tilde{\chi}-\tilde{\phi}_{n})/\sqrt{2}})^T= U(x) \psi(x)$, the unitary matrix $U(x)=T_x \exp[-i\int_{0}^{x}dx' (\xi(x')\sigma^{+}+\xi^{*}(x')\sigma^{-})/v_n]$ represents
random disorder scattering, $\psi\equiv (e^{i(\chi+\phi_{n})/\sqrt{2}}, e^{i(\chi-\phi_{n})/\sqrt{2}})^T$ is a two-component fermionic operator, $\chi$ is an auxiliary bosonic field, and 
$\sigma^{\pm}=\sigma_{x}\pm i\sigma_{y}$, $\sigma_{x}$ and $\sigma_{y}$ are the Pauli matrices.
The equal-position correlator $\langle [\Psi_{\pm}^{\dagger}(x_{\mathcal{L}},t), \Psi_{\pm}(x_{\mathcal{L}},0)] \rangle$
is replaced by $\langle [\tilde{\Psi}_{\pm}^{\dagger}(x_{\mathcal{L}},t), \tilde{\Psi}_{\pm}(x_{\mathcal{L}},0)] \rangle$ when
we choose the global gauge transformation making $U(x_\mathcal{L}) = 1$.
Then, it is readily computed because the Hamiltonian is free in the basis of $\tilde{\Psi}_{\pm}$, and is the same as $G_{j} (x_\alpha, x_\alpha; t)$ in Eq.~\eqref{conductance11} that is obtained in the absence of interaction ($v=0$) between the charge
and neutral components and disorder ($W=0$). Hence,  Eqs.~\eqref{conductancevcsimvn}$-$\eqref{conductancevcggvn2} can be also applied to the strong disorder regime.

\section{Discussion and conclusion}

We have studied electron dephasing at $\nu = 2/3$.
Electron fractionalization  into charge and neutral components 
leads to plasmonic dephasing. 
When $v_c \gg v_n$ (which is likely \cite{Wan,Lee}) and at $k_B T > E_n/(4\pi^2 \delta_n)$, a new type of dephasing additionally arises.
This dephasing is topological, resulting from
the fractionalization and the fractional braiding statistics of the components, and occurs depending on the topological sectors characterized by the winding numbers $(\delta p, \delta q)$ of the components; its dependence on the even-odd parity of $\delta p + \delta q$ is mathematical reminiscent of
the parity (integer versus half-integer spin) dependent role of
the topological $\theta$ term in antiferromagnetic spin chains~\cite{Haldane2}. 
It leads to period halving of the AB oscillations.


We emphasize that the topological dephasing occurs in both the regimes of strong and weak disorder, when bulk-edge interactions are not strong. In the case of weak disorder,  which may be realized in high temperatures, 
weak intermode interaction causes the decay of the plasmonic part of the neutral component, accompanied by the diffusive spreading of the plasmonic part of the charge component~\cite{Kane_PRB}.
These do not affect  the  topological dephasing, hence the emergence of the $h/2e$ 
oscillations. On the other hand, in the case of strong disorder at low temperatures, $v$ renormalizes towards zero~\cite{Kane_PRL}, and then does not change the physics of the topological dephasing.
Note that bulk-edge Coulomb interactions become weaker in QDs of larger area; the pure AB regime (or the intermediate regime between the pure AB and the Coulomb-dominated regimes) could be achieved when the edge-to-bulk capacitance is smaller than other capacitances even when strong backscattering occurs at QPCs.



We also note that the QH edges at $\nu = 2/3$ may undergo more complex edge reconstruction at about $T > 50$ mK~\cite{Neutral.Wang,Neutral.Bid2}. At lower temperature our analysis is applicable, while at higher temperature different topological dephasing may occur. 
Assuming $v_n \sim 5 \times 10^4$~m/s, $v_c \sim 5 \times 10^5$~m/s, and $L=10 \, \mu \textrm{m}$, we expect that the $h/(2e)$ oscillation will appear at 
temperature $k_B T > \hbar v_n / (2\pi \delta_n L ) \sim 20$ mK.
In this case, the oscillation will be suppressed at $k_B T > \hbar v_c / (2\pi \delta_c L ) \sim 60$ mK, due to the plasmonic dephasing.

Detection of the period halving supports the topological dephasing, thus,
the fractional statistics of the charge component at $\nu = 2/3$. 
The plasmonic dephasing and the topological dephasing will occur, with modifications, in other anyon interferometers or at other $\nu$'s.

It should be mentioned that the known mechanisms yielding $h/(2e)$ oscillations in other
mesoscopic systems do not apply to our setup.
The Altshuler-Aronov-Spivak mechanism~\cite{Altshuler, Murat} employs disorder averaging in multi-channel geometries, which is not present in our setup.
Another mechanism for $h/2e$ oscillations~\cite{Kataoka, Sim} relies on integer QH edge modes in an antidot at temperatures much below the charging energy of the antidot.
Moreover, our setup does not have superconducting fluctuations that support such periodicity.

\begin{acknowledgements}

We thank G. Rafael for a helpful discussion. HSS was supported by Korea NRF (Grant No. NRF-2010-00491 and Grant No. NRF-2013R1A2A2A01007327). 
YG was supported 
by DFG (Grant No. RO 2247/8-1).

\end{acknowledgements}

\appendix

\section{Quantum-Dot Hamiltonian}
\label{Appen_QD}


We derive the Hamiltonian $H_\textrm{D}$ (cf. Eq.~\eqref{Hamiltoniandot} in the main text).  $H_\textrm{D}$ is written in terms of 
the mode $\phi_1$ of filling factor $\nu_1=1$ and the counterpropagating mode $\phi_2$ of $\nu_2=-1/3$, as
\begin{align} \label{HamiltonianDotOrg_sup} 
H_{\textrm{D}}&=\frac{\hbar}{4\pi}\int_{0}^{L} dx [v_1 (\partial _x \phi_1)^2 + 3 v_2 (\partial _x \phi_2)^2 + 2v_{12}\partial_x \phi_1 \partial_x \phi_2] \nonumber \\
&+\int_{0}^{L} dx [\xi (x) \exp(i\phi_1+3i\phi_2)+\text{H.c.}].
\end{align}
%
$v_{1(2)}$ is the velocity of $\phi_{1(2)}$ (renormalized by the intra-mode interactions) 
and $v_{12}$ describes the inter-mode interaction.
$\phi_{1,2}$ satisfies 
$[\phi_i(x),\phi_{i'}(x')]=i\pi \nu_i^{\textrm{edge}} \text{sgn}(x-x')\delta_{ii'}$.
Each field $\phi_i(x)$ is decomposed through 
$\phi_i(x)=\phi^{\text{pl}}_{i}(x) + \phi^{0}_{i}(x)$ into a plasmonic mode  $\phi^{\text{pl}}_{i}(x)$, satisfying the periodic boundary condition of $\phi^{\text{pl}}_{i}(x+L) = \phi^{\text{pl}}_{i}(x)$, and a zero mode $\phi^{0}_{i}(x)$,
\begin{equation} 
\phi_i^{0}(x) = \frac{2\pi \nu_i x}{L} (N_i + \frac{1}{2} - \frac{\Phi}{\Phi_0}) - \lambda_i. 
\label{zero_12} 
\end{equation}
The number operator $N_i$ counts the excess number of quasiparticles of charge $\nu_i^{\textrm{edge}}e$. Its canonical conjugate $\lambda_i$ satisfies $[\lambda_i, N_{i'}] = i \delta_{i i'}$,
and $e^{\pm i \lambda_{i}}$ changes $N_i$ by $\pm 1$, acting as a Klein factor.
This ensures $[\phi_i^{0}(x), \phi_{i'}^{0}(x')] =2 i \delta_{ii'} \pi \nu_i^{\textrm{edge}} (x-x')/L $.
Combined with the commutation rule of the plasmonic part $[\phi_i^{\text{pl}}(x), \phi_{i'}^{\text{pl}}(x')] =i\pi \nu_i^{\textrm{edge}} (\text{sgn}(x-x')-2 (x-x')/L)\delta_{ii'}$,
this leads to $[\phi_i(x),\phi_{i'}(x')]=i\pi \nu_i^{\textrm{edge}} \text{sgn}(x-x')\delta_{ii'}$.
The term $1/2$ in the bracket of Eq.~\eqref{zero_12} is introduced, to impose the boundary condition of electron operators, $\exp({i \phi_i(x+L) / \nu_i^{\textrm{edge}}}) = \exp({i \phi_i(x) / \nu_i^{\textrm{edge}}})\exp({-2\pi i \Phi/\Phi_0})$.  The magnetic flux $\Phi$ enclosed by the QD edge states causes the shift of $N_i \rightarrow N_i - \Phi/\Phi_0$ in Eq.~\eqref{zero_12}; as $\Phi$ increases (decreases) by $\Phi_0$, the edge state with filling factor $\nu_i^{\textrm{edge}}$ is energetically stabilized by removing (adding) its own quasiparticle of charge $\nu_i^{\textrm{edge}} e$.

Combining $\phi_i$'s, one introduces the charge mode $\phi_c = \sqrt {3/2} (\phi_1 + \phi_2)$ and the neutral mode $\phi_n = \sqrt {1/2} (\phi_1 + 3\phi_2)$,
satisfying
$[\phi_{c/n}(x), \phi_{c/n}(x')]=\pm i\pi \text{sgn}(x-x')$ and $[\phi_c(x), \phi_n(x')]=0$.
Putting this into Eq.~\eqref{HamiltonianDotOrg_sup},
we derives Eq.~(2).

%

\section{Derivation of $N_c$ and $N_n$}
\label{AppenNumber}

We derive Eq.~\eqref{OldnumberNewnumber} in the absence of disorder ($W=0$). The charge (neutral) mode is decomposed into the zero-mode part $\phi_c^{0}(x)$ ($\phi_n^{0}(x)$) and the plasmonic part $\phi_c^\textrm{pl}(x)$ ($\phi_n^\textrm{pl}(x)$). The latter describes edge plasmonic excitations, while the former anyon number excitations. 
The zero-mode parts are determined from Eq.~\eqref{zero_12} as 
\begin{align} \label{zeromode_sup}
\phi_{c}^{0}(x)&=\sqrt{\frac{1}{6}}\frac{2\pi x}{L}\bigg(N_{c}+1 - 2\frac{\Phi}{\Phi_0} \bigg)-\sqrt{6}\lambda_{c}, \nonumber \\ 
\phi_{n}^{0}(x)&=-\sqrt{\frac{1}{2}}\frac{2\pi N_{n}  x}{L}-\sqrt{2}\lambda_{n}.
\end{align}
We impose $[\lambda_{c(n)}, N_{c(n)}] = i$, $[\lambda_{c(n)}, N_{n(c)}]=0$, and $[\phi_{c(n)}^{\text{pl}}(x),\phi_{c(n)}^{\text{pl}}(x')]=\pm i \pi \text{sgn}(x-x') \mp 2\pi i (x-x')/L$. 
These ensure $[\phi_{c/n}^{0}(x),\phi_{c/n}^{0}(x')]=\pm 2\pi i (x-x')/L$  and  $[\phi_{c/n}(x), \phi_{c/n}(x')]=\pm i\pi \text{sgn}(x-x')$. 
Comparing Eqs.~\eqref{zeromode_sup} with \eqref{zero_12}, one gets Eq.~\eqref{OldnumberNewnumber}. 

\section{Derivation of differential conductance}
\label{secAppendix_seq}

In this appendix, we derive $\mathcal{G}$ in Eq.~\eqref{conductance11}.
The electron current $\mathcal{I}_{\mathcal{R}}$ along the right lead edge is given by 
$\mathcal{I}_{\mathcal{R}}$,
\begin{widetext}
\begin{align} \label{electroncurrent}
\langle \mathcal{I}_{\mathcal{R}} \rangle &=  e\frac {d \langle n_{\mathcal{R}} \rangle}{dt} = i\frac{e}{\hbar} \langle [H_\textrm{T}, n_\mathcal{R}] \rangle 
\nonumber \\
&=-\frac{ie}{\hbar} \sum_{i,j=\pm} \bigg \langle \textrm{T}_t \exp \bigg(\frac{i}{\hbar} \int_{-\infty}^{0} dt\hat{H}_{\textrm{T}} (t)\bigg) \big [t_{\mathcal{R}ij} \hat{\Psi}_{\mathcal{R},i}^{\dagger}(0,t'=0) \hat{\Psi}_{j}(x_{\mathcal{R}},t'=0)-\text{H.c.}\big ] \textrm{T}_t \exp \bigg(-\frac{i}{\hbar} \int_{-\infty}^{0} dt \hat{H}_{\textrm{T}} (t)\bigg) \bigg \rangle, 
\end{align}
\end{widetext}
$n_{\mathcal{R}}$ counts electron number in the right lead edge, 
$\textrm{T}_t$ is the time ordering, the operators with (without) caret are in the interaction (Heisenberg) picture, 
and $\langle \cdot \rangle$ is the thermal  average. 
We set $t=0$ at which $\mathcal{I}_{\mathcal{R}}$ is measured.
To second order in the tunneling strengths,
$\langle \mathcal{I}_\mathcal{R} \rangle$ is calculated as 
\begin{widetext}
\begin{align}
 \langle \mathcal{I}_{\mathcal{R}}^{(2)} \rangle & = \frac{e}{\hbar^2} \sum_{i,j=\pm}\int_{-\infty}^{0} dt\big\langle \big[\hat{H}_\textrm{T} (t), (t_{\mathcal{R}ij} \langle \hat{\Psi}
_{\mathcal{R},i}^{\dagger}(0,0) \hat{\Psi}_{j}(x_{\mathcal{R}},0)\rangle - \text{H.c.})\big]
\big\rangle \nonumber \\ &= \frac{e}{\hbar^2} \sum_{i,j=\pm} |t_{\mathcal{R}ij}|^2 \int_{-\infty}^{0} dt\text{Re} \Big [\exp \big(-\frac{i}{\hbar}(\mu_\mathcal{R}-\mu_\textrm{D})t\big ) \Big (  G_{\textrm{D}, j}^K(-t) (G_{\mathcal{R}, i}^R-G_{\mathcal{R}, i}^A)(t)- 
  (G_{\textrm{D}, j}^R-G_{\textrm{D}, j}^A)(-t) G_{\mathcal{R}, i}^K(t)\Big) \Big ]. 
\end{align}
$G_{\alpha (\textrm{D}), \pm}^{R}$, $G_{\alpha (\textrm{D}), \pm}^{A}$ and $G_{\alpha (\textrm{D}), \pm}^{K}$ are the retarded, advanced, and Keldysh Green's functions of lead edge $\alpha =\mathcal{L}, \mathcal{R}$ (QD), 
\begin{align} \label{green'sfunction}
(G_{\alpha, \pm}^R-G_{\alpha, \pm}^A)(t) &\equiv -i\langle \{ \hat{\Psi}_{\alpha, \pm}(0,t), \hat{\Psi}_{\alpha, \pm}^{\dagger}(0,0)\}\rangle, 
\,\,\,\,\,\,\,\,\,\,\,\,\,\,
G_{\alpha, \pm}^K(t)\equiv -i\langle [ \hat{\Psi}_{\alpha, \pm}(0,t), \hat{\Psi}_{\alpha, \pm}^{\dagger}(0,0)]\rangle,\nonumber\\
(G_{\textrm{D}, \pm}^R-G_{\textrm{D}, \pm}^A)(t)&\equiv -i\langle \{ \hat{\Psi}_{\pm}(x_{\alpha},t), \hat{\Psi}_{\pm}^{\dagger}(x_{\alpha},0)\}\rangle, 
\,\,\,\,\,\,\,\,\,\,\,\,\,\,
G_{\textrm{D}, \pm}^K(t)\equiv -i\langle [ \hat{\Psi}_{\pm}(x_{\alpha},t), \hat{\Psi}_{\pm}^{\dagger}(x_{\alpha},0)]\rangle, 
\end{align}
and $\mu_{\alpha}$ ($\mu_\textrm{D}$) is the chemical potential for lead edge $\alpha$ (the QD). $\mu_\textrm{D}$ is assumed to be uniform over the entire region of the QD, which is valid in the linear response regime. 
The expression $\langle \mathcal{I}_\mathcal{L}^{(2)} \rangle$ of electron current in the left lead edge is similar to that of $\langle \mathcal{I}_\mathcal{R}^{(2)} \rangle$.
To second order in the tunneling strengths, the current $\mathcal{I}$ through the QD is written as $\mathcal{I} = - \langle \mathcal{I}_\mathcal{L}^{(2)} \rangle = \langle \mathcal{I}_\mathcal{R}^{(2)} \rangle$.
\end{widetext}

Applying the current conservation condition of 
$\langle \mathcal{I}_{\mathcal{R}}^{(2)} \rangle +\langle \mathcal{I}_{\mathcal{L}}^{(2)} \rangle =0$, we write the symmetrized form of $\mathcal{I}$ as
\begin{widetext}
\begin{align} 
\mathcal{I}&=\frac{\gamma_{\mathcal{L} } \langle \mathcal{I}_{\mathcal{R}}^{(2)} \rangle- \gamma_{\mathcal{R}} \langle \mathcal{I}_{\mathcal{L}}^{(2)} \rangle}{ \gamma_{\mathcal{L}}+\gamma_{\mathcal{R}}}\,\,\,\,\,\,\,\,\,\,
\big[\gamma_{\mathcal{L}/\mathcal{R}} = \sum_{i,j=\pm}|t_{\mathcal{R}/\mathcal{L}ij}|^2/ (\hbar a v_c^{3/4} v_n^{1/4})\big]\nonumber \\ 
	&= \frac{e}{4\hbar} a v_c^{3/4} v_n^{1/4}  \frac{ \gamma_{\mathcal{L}}\gamma_{\mathcal{R}}}{ \gamma_{\mathcal{L} }+\gamma_{\mathcal{R}}} \sum_{i,j=\pm}\int_{-\infty}^{0} dt \nonumber \\ 
 & \times 
 \text{Re} \Big [ \big (e^{-\frac{i}{\hbar}(\mu_{\mathcal{R}}-\mu_{\textrm{D}})t}-e^{-\frac{i}{\hbar}(\mu_{\mathcal{L}}-\mu_{\textrm{D}})t} \big )
\Big ( G_{\textrm{D}, j}^K(-t) (G_{\mathcal{R}, i}^R-G_{\mathcal{R}, i}^A)(t)- 
  (G_{\textrm{D}, j}^R-G_{\textrm{D}, j}^A)(-t) G_{\mathcal{R}, i}^K(t)\Big) \Big].
\end{align}
\end{widetext}
In the second equality, we used the simplification that
the left and right lead edges are symmetric ($H_\mathcal{L} = H_\mathcal{R}$), namely
$G^{R}_{\mathcal{L}, \pm} = G^{R}_{\mathcal{R}, \pm}$,
$G^{A}_{\mathcal{L}, \pm} = G^{A}_{\mathcal{R}, \pm}$, and
$G^{K}_{\mathcal{L}, \pm} = G^{K}_{\mathcal{R}, \pm}$.
We also used the fact that the Green's functions are independent 
of the index $i = \pm$ of the electron field operators in $H_\textrm{T}$. 
The differential
conductance $\mathcal{G} = d \mathcal{I}/ d V |_{V \to 0}$ is written as
\begin{widetext}
\begin{align} \label{currentAndconductance}
\mathcal{G} = -\frac{e^2}{8\hbar^2} a v_c^{3/4} v_n^{1/4}  \frac{ \gamma_{\mathcal{L}}\gamma_{\mathcal{R}}}{ \gamma_{\mathcal{L} }+\gamma_{\mathcal{R}}} \sum_{i,j} \sum_{\alpha=\mathcal{R},\mathcal{L}} \int_{-\infty}^{0} dt 
\text{Im} \Big [ t \Big ( G_{D, j}^K(-t) (G_{\alpha, i}^R-G_{\alpha, i}^A)(t)- (G_{D, j}^R-G_{D, j}^A)(-t) G_{\alpha, i}^K(t)\Big)\Big ], \nonumber
\end{align}
\end{widetext}
where $eV \equiv \mu_\mathcal{L} - \mu_\mathcal{R}$.
The Green's functions for the lead edges are computed as 
\begin{widetext}
\begin{align}
(G_{\alpha, i}^R-G_{\alpha, i}^A)(t) &= -\frac{i}{\pi a} \text{Re} \bigg[ \bigg (\frac{\sinh(\frac{i\pi a k_B T}{\hbar v_{c}})}{\sinh \big(\frac{\pi k_B T}{\hbar v_{c}}(ia-v_{c}t)\big)} \bigg )^{\frac{3}{2}}
\bigg (\frac{\sinh(\frac{i\pi a k_B T}{\hbar v_{n}})}{\sinh \big(\frac{\pi k_B T}{\hbar v_{n}}(ia-v_{n}t)\big)} \bigg )^{\frac{1}{2}}  \bigg], \nonumber \\
G_{\alpha, i}^K(t) &= \frac{1}{\pi a} \text{Im} \bigg[ \bigg (\frac{\sinh(\frac{i\pi a k_B T}{\hbar v_{c}})}{\sinh \big(\frac{\pi k_B T}{\hbar v_{c}}(ia-v_{c}t)\big)} \bigg )^{\frac{3}{2}}
\bigg (\frac{\sinh(\frac{i\pi a k_B T}{\hbar v_{n}})}{\sinh \big(\frac{\pi k_B T}{\hbar v_{n}}(ia-v_{n}t)\big)} \bigg )^{\frac{1}{2}}  \bigg].
\end{align}
\end{widetext}
The Green's functions are independent of the index $\alpha=\mathcal{L},\mathcal{R}$, because of the imposed symmetry between the left and right lead edges. Since $(G_{\alpha, i}^R-G_{\alpha, i}^A)(t)  \gg G_{\alpha, i}^K(t)$ at $|t| > a/ v_n$, and since the processes of $|t| < a / v_n$ do not contribute to the interference in $\mathcal{G}$, we can ignore $G_{\alpha, i}^K(t)$ in the expression of $\mathcal{G}$. Then, a  simplified form of $\mathcal{G}$ is obtained as
\begin{widetext}
\begin{equation} 
\mathcal{G} = \frac{e^2}{4\hbar^2}  \frac{\gamma_{\mathcal{L}}\gamma_{\mathcal{R}}}{ \gamma_{\mathcal{L}}+\gamma_{\mathcal{R}}} \frac{a^2 k_B T}{\hbar v_c^{3/4} v_n^{1/4}} \sum_{j=\pm} \sum_{\alpha=\mathcal{R},\mathcal{L}} \int_{-\infty}^{0} dt F(t)\text{Im} G_j(x_{\alpha}, x_{\alpha}; t)=\frac{e^2}{\hbar}  c_g \tilde{\gamma} k_B T  \sum_{j=\pm} \sum_{\alpha=\mathcal{R},\mathcal{L}} \int_{-\infty}^{0} dt F(t)\text{Im} \, G_j(x_\alpha, x_\alpha; t).  \nonumber
\end{equation} 
\end{widetext}
Here, $\tilde{\gamma} = \gamma_{\mathcal{L}}\gamma_{\mathcal{R}}/( \gamma_{\mathcal{L}}+\gamma_{\mathcal{R}})$, $c_g = a^2 / (4 \hbar^2 v_c^{3/4} v_n^{1/4})$, the weight factor $F(t) = (\pi k_B T t / \hbar) \sinh^{-2}(\pi k_B T t / \hbar)$,  
and
$G_j(x_{\alpha}, x_{\alpha}; t) \equiv \big\langle \big [\Psi_{j}^{\dagger}(x_{\alpha},t), \Psi_{j}(x_{\alpha},0) \big ]
\big\rangle$  represents a Green's function of the QD; its starting position $x_\alpha$ is the same with the ending one in the sequential tunneling regime. 
The weight factor $F(t)$ decays rapidly as $e^{-2\pi k_B T t / \hbar}$ for $t \gg \hbar / k_B T$, describing the plasmonic dephasing (by partial overlap due to different positions of the neutral components between the interfering paths) occuring at lead edge $\alpha$. 
The above is the derivation of Eq.~\eqref{conductance11} in the main text.

Below, we further compute $\mathcal{G}$ in the case of $v=0$.
We note that in the absence of disorders ($W=0$) and interaction ($v=0$) between the two modes, the Hamiltonian $H_{\textrm{D}}^{0}$ of the zero-mode parts is obtained
\begin{align} \label{HamiltonianZero_sup}
H_{\textrm{D}}^{0} &= \frac{\pi \hbar v_c}{6L}\bigg(N_{c}+1-2\frac{\Phi}{\Phi_0}\bigg)^2 +
\frac{\pi \hbar v_n}{2L} N_{n}^2  \nonumber \\
&= \frac{E_c}{12} \bigg(N_{c}+1-2\frac{\Phi}{\Phi_0}\bigg)^2 + \frac{E_n}{4}  N_{n}^2.
\end{align}
Here we define energy scales  $E_{c(n)} \equiv 2\pi \hbar v_{c(n)}/ L$.
Then, the Green's function $G_j$ is decomposed into the charge and neutral components. Then $\mathcal{G}$ is simplified as
\begin{align} \label{Sup:conductance}
\mathcal{G}
& =\frac{e^2}{ h} \frac{\tilde{\gamma}  a k_B T}{\hbar^2 v_c^{3/4} v_n^{1/4}}  \int_{-\infty}^{\infty} dt    F_{\alpha} (t)  \text{Re} G^{0}(t) \nonumber \\ & \times \text{Im} \big[G_c(t) 
G_n(t) e^{-3\pi i v_c t/2L} e^{-\pi i v_n t/2L} \big] , 
\end{align}
where the plasmonic parts $G_{c}(t)$ and $G_n (t)$ of the charge and neutral modes and the zero-mode part $G^{0}(t)$ are 
\begin{align}
G_c(t) &= \langle e^{i\sqrt{\frac{3}{2}} \phi_c(t)} e^{-i\sqrt{\frac{3}{2}} \phi_c(0)}\rangle \nonumber \\&=  \bigg(\exp\big({\frac{i\pi v_c t}{L}}\big)\frac{ \theta_1 (-\frac{i\pi a}{L}
, e^{-\gamma_{c}})}{\theta_1 (\frac{\pi (v_{c}t -ia)}{L}
, e^{-\gamma_{c}})} \bigg )^{\frac{3}{2}}, \nonumber \\ 
G_n(t) &= \langle e^{i\sqrt{\frac{1}{2}} \phi_n(t)} e^{-i\sqrt{\frac{1}{2}} \phi_n(0)}\rangle \nonumber \\& = \bigg ( \exp\big({\frac{i \pi v_n t}{L}}\big) \frac{ \theta_1 (-\frac{i\pi a}{L}
, e^{-\gamma_{n}})}{\theta_1 (\frac{\pi (v_{n}t-ia)}{L}
, e^{-\gamma_{n}})} \bigg )^{\frac{1}{2}},  \nonumber \\
G^{0} (t) &=\big \langle e^{i \pi v_{c} t (N_{c}+1-2\Phi/\Phi_0 )/L}  e^{-i \pi N_{n} v_{n} t /L} \big \rangle. 
\label{Greenplasmonic_sup}
\end{align}
The elliptic-theta function of the first kind is $\theta_1 (z, q) = 2 q^{1/4} \sin z \prod\limits_{n=1}^{\infty} (1-2q^{2n} \cos(2z) + q^{4n})(1-q^{2n})$, and $\gamma_{c(n)} \equiv \pi \hbar v_{c(n)} / (k_B T L) $ \cite{Eggert}. 
In the derivation of Eq.~\eqref{Sup:conductance}, we used the relations of
$G_{j=+} (x_\alpha, x_\alpha, t) = G_{j=-} (x_\alpha, x_\alpha, t)$, 
$G_{c(n)}(-t) = G_{c(n)}^{*}(t)$, and $F(t) = -F(-t)$. The zero-mode part $G^{0}(t)$ will be calculated in Appendix~\ref{appendix_top}.

\section{Topological dephasing}
\label{appendix_top}

We first sketch the topological dephasing in the case of $v_c \gg v_n$ and no disorder, and derive it, by expanding the zero-mode contribution to $\mathcal{G}$ in harmonics of the winding numbers. 
The discussion is valid even with strong disorder.

As in the main text, we consider the interference between the processes in Figs.~2(a) and 
2(c).
In the semiclassical regime of $L \gg L_{T,c/n}$, counting the winding numbers $p$ and $q$ of the {\it center} of the spatial distributions of the charge and neutral components, we find that
this interference contributes mainly to $(\delta p, \delta q) = (1,0)$, and gains the net phase of $\theta =\pi (N_c - 2\Phi / \Phi_0)\delta p$ from those windings that braid with $N_c$ charge excitations and $N_n$ neutral excitations.
At lower temperature of $L \gtrsim  L_{T,c/n}$, the tails (namely the spatial width $L_{T,c/n}$) of the spatial distributions of the components are non-negligible, and imply that the two processes can also contribute to quantum mechanical net windings $(\delta p_{\textrm{qm}}, \delta q_{\textrm{qm}})$ that can differ from $(\delta p, \delta q)$. To see the contribution to different windings, we expand the average of $\langle e^{i\theta} \rangle_{k_B T}$ over the thermal fluctuations of $N_c$ and $N_n$ in the harmonics of $(\delta p_{\textrm{qm}}, \delta q_{\textrm{qm}})$,
\begin{align}
\langle e^{i\theta} \rangle_{k_B T} &= \sum_{\delta p_{\textrm{qm}}, \delta q_{\textrm{qm}} \in \mathbb{Z}} f(\delta p_{\textrm{qm}}, \delta q_{\textrm{qm}}, k_B T) 
\nonumber \\ &\times \exp (2\pi i  \delta p_{\textrm{qm}} \Phi / \Phi_0) \exp (2\pi i \delta q_{\textrm{qm}} \varphi_n ), \end{align}
where we introduce a fictitious "neutral flux" $\varphi_n$ in the mathematical analogy of $\Phi/\Phi_0$ in order to have the expansion; $\theta$ is now generalized to $\theta =\pi (N_c - 2\Phi / \Phi_0)\delta p + \pi (N_n - 2 \varphi_n)\delta q$, and we put  $\varphi_n \rightarrow 0$ at the end.
The thermal fluctuations of $N_c$ and $N_n$ are governed by 
the QD-energy $H_{\textrm{D}}^{0}=E_c (N_c - 2\Phi / \Phi_0 + 1)^2 / 12 + E_n (N_n-2\varphi_n)^2 / 4$ (cf. Eq.~\eqref{HamiltonianZero_sup}). 
Notice $H_{\textrm{D}}^{0}(\Phi/\Phi_0, \varphi_n) |_{N_c, N_n} = H_{\textrm{D}}^{0}(\Phi/\Phi_0 +1, \varphi_n +1) |_{N_c + 2, N_n +2}$ and $\theta(\Phi / \Phi_0)|_{N_c, N_n} = \theta(\Phi / \Phi_0 + 1)  |_{N_c + 2, N_n +2}$, meaning that
 $\theta$ and  $H^0_\textrm{D}$ are restored by changing $N_c$ and $N_n$ by 2, when each flux shifts by one as $\Phi/\Phi_0 \rightarrow \Phi/\Phi_0 +1$ and $\varphi_n \rightarrow \varphi_n+1$. 

We decompose the amplitude $f(\delta p_{\textrm{qm}}, \delta q_{\textrm{qm}}, k_B T) = f_\textrm{e} (\delta p_{\textrm{qm}}, \delta q_{\textrm{qm}}, k_B T) + f_\textrm{o} (\delta p_{\textrm{qm}}, \delta q_{\textrm{qm}}, k_B T)$ into the average $f_\textrm{e}$ over (even $N_c$, even $N_n$) and that $f_\textrm{o}$ over (odd $N_c$, odd $N_n$); $N_c$ and $N_n$ should have the same parity, according to Eq.~(3) in the main text. 
We find a useful relation of $f_\textrm{o} (\delta p_{\textrm{qm}}, \delta q_{\textrm{qm}}, k_B T) = (-1)^{\delta p_{\textrm{qm}}+\delta q_{\textrm{qm}}} f_\textrm{e} (\delta p_{\textrm{qm}}, \delta q_{\textrm{qm}}, k_B T)$, obtained from the fact that 
the thermal average of $\langle e^{i\theta} \rangle_{k_B T}$ over odd $N_c$ and odd $N_n$ 
at $(\Phi / \Phi_0, \varphi_n)$ is identical to 
that over even $N_c$ and even $N_n$ at $(\Phi / \Phi_0+1/2, \varphi_n+1/2)$, according to $H^0_\textrm{D}$.
This relation leads to $f(\delta p_{\textrm{qm}}, \delta q_{\textrm{qm}}, k_B T) = 0$ for odd $\delta p_{\textrm{qm}}+ \delta q_{\textrm{qm}}$, describing the topological dephasing.
For the interference  between those in Figs.~2(a) and 2(c), the contribution from the semiclassical winding of $(\delta p, \delta q) = (\delta p_{\textrm{qm}}, \delta q_{\textrm{qm}}) = (1,0)$ vanishes (independent of temperature $T$!), while $\mathcal{G}_{\delta p =1}$ is contributed dominantly from
the quantum mechanical winding of  $(\delta p_{\textrm{qm}}, \delta q_{\textrm{qm}}) = (1,-1)$  of the tail.



We confirm the above discussion mathematically.
The flux dependence of $\mathcal{G}$ in Eq.~\eqref{g_decomp} comes from the zero-mode part of the electron field operator $\Psi_\pm$, hence, from the Green's function $G^0$ in Eq.~\eqref{Greenplasmonic_sup}. 
Using Eq.~\eqref{HamiltonianZero_sup}, $G^0 = \langle \exp [i \pi v_{c}t (N_{c}- 2 \Phi/\Phi_0+1) / L ]
\exp(-i\pi  v_{n}t N_{n} / L) 
\rangle$ is computed as 
\begin{widetext}
\begin{align}
G^{0}(t)
& =  
\Big [ \sum\limits_{n_c, n_n =-\infty}^{\infty} 
\Big \{ \exp \Big (- \frac{\frac{E_c}{12} \big(2n_c + 1 -\frac{2\Phi}{\Phi_0} + 1\big)^2 +  \frac{E_n}{4} \big(2n_n + 1\big)^2}{k_B T} \Big)
e^{i\frac{\pi  v_c t}{L} (2 n_c + 1 -\frac{ 2 \Phi}{\Phi_0} +1 )} 
e^{-i\frac{\pi v_n t}{L} (2 n_n + 1)}
\nonumber \\ &
+\exp \Big (- \frac{\frac{E_c}{12} \big(2n_c-\frac{2\Phi}{\Phi_0}+1\big)^2 +  \frac{E_n}{4} (2n_n)^2}{k_B T} \Big)
e^{i \frac{\pi v_c t}{L} (2n_c -\frac{2\Phi}{\Phi_0}+1)}
e^{-i \frac{\pi v_n t}{L} (2n_n)} 
\Big \} \Big] \nonumber \\ &
\Big /
 \Big [ \sum\limits_{n_c, n_n =-\infty}^{\infty} 
\Big \{ \exp \Big (-\frac{\frac{E_c}{12} \big(2n_c + 1 -\frac{2\Phi}{\Phi_0} +1\big)^2 +  \frac{E_n}{4} \big(2n_n + 1\big)^2}{k_B T} \Big) +\exp \Big (- \frac{\frac{E_c}{12} \big(2n_c -\frac{2\Phi}{\Phi_0} +1\big)^2 +  \frac{E_n}{4} (2n_n)^2}{k_B T} \Big) \Big \} \Big ].
\label{Zeromodeaverage_sup}  
\end{align}
\end{widetext}
The first term of Eq.~\eqref{Zeromodeaverage_sup} comes from odd integers $N_c=2n_c +1$ and $N_n=2n_n +1$, and the second from even integers $N_c = 2n_c$ and $N_n = 2n_n$.
Utilizing the Poisson summation formula of $\sum_{n=-\infty}^{\infty} \exp [-a(n+\delta)^2] \exp [2bi(n+\delta) ] =\sum_{p' = -\infty}^{\infty} \exp(-2\pi i p' \delta) \exp [-(\pi p' +b)^2/a]$ with real constants, $a$, $b$ and $\delta$, we obtain
\begin{widetext}
\begin{align}
G^{0}(t)
& =  \frac{\sum\limits_{\delta p', \delta q' = -\infty}^{\infty} (-1)^{\delta p'} (1+(-1)^{\delta p'+\delta q'})\exp(2\pi i \delta p' \Phi/ \Phi_0)
\exp \big [-\frac{3 \pi^2 k_B T}{E_c}\big(\frac{v_{c} t}{L} - \delta p' \big)^2 \big] \exp \big[-\frac{\pi^2 k_B T}{E_n}\big( \frac{v_{n} t}{L} +\delta q'\big)^2 \big]}
{\sum\limits_{\delta p'', \delta q'' = -\infty}^{\infty}
(-1)^{\delta p''} (1+(-1)^{\delta p''+\delta q''})\exp (2\pi i \delta p'' \Phi/ \Phi_0)\exp \big(-\frac{3 \pi^2 k_B T}{E_c}{\delta p''}^2 \big ) \exp \big (-\frac{\pi^2 k_B T}{E_n}{\delta q''}^2 \big )} \nonumber \\ &=
\frac{\sum\limits_{\delta p'+ \delta q' \in 2 \mathbb{Z}} (-1)^{\delta p'} 
\exp(2\pi i \delta p'\Phi/ \Phi_0)
\exp \big [-\frac{3 \pi^2 k_B T}{E_c}\big(\frac{v_{c} t}{L} - \delta p' \big)^2 \big] \exp \big[-\frac{\pi^2 k_B T}{E_n}\big( \frac{v_{n} t}{L} +\delta q'\big)^2 \big]}
{\sum\limits_{\delta p''+ \delta q''\in 2 \mathbb{Z}}
(-1)^{\delta p''} \exp (2\pi i \delta p'' \Phi/ \Phi_0)\exp \big(-\frac{3 \pi^2 k_B T}{E_c}{\delta p''}^2 \big ) \exp \big (-\frac{\pi^2 k_B T}{E_n}{\delta q''}^2 \big )}. 
\label{zeromodeaverage2_sup}
 \end{align}
\end{widetext}
Here $2\mathbb{Z}$ is the set of even integers. The denominator is approximated as 1 at high temperature $k_B T \gg \hbar v_c /L$; this condition of $k_B T \gg \hbar v_c /L$ is chosen for simplicity, and it is not a condition for the topological dephasing.
Then $G^{0}(t)$ is written by the harmonics of winding numbers $\delta p' \to \delta p_\textrm{qm}$ and $\delta q' \to \delta q_\textrm{qm}$,
\begin{align}
G^{0}(t)&\simeq
\sum\limits_{\delta p_{\textrm{qm}}+ \delta q_{\textrm{qm}} \in 2 \mathbb{Z}} (-1)^{\delta p_{\textrm{qm}}} 
\exp(2\pi i \delta p_{\textrm{qm}}\Phi/ \Phi_0) \nonumber \\ & \times
\exp \big [-\frac{3 \pi^2 k_B T}{E_c}\big(\frac{v_{c} t}{L} - \delta p_{\textrm{qm}} \big)^2 \big] 
\nonumber \\ & \times
\exp \big[-\frac{\pi^2 k_B T}{E_n}\big( \frac{v_{n} t}{L} +\delta q_{\textrm{qm}}\big)^2 \big]. \label{G0}
\end{align}
Notice that the windings of odd $\delta p_{\textrm{qm}} + \delta q_{\textrm{qm}}$ do not contribute to $G^{0}(t)$, as mentioned in the main text.

\section{Semiclassical approximation: Derivation of  Eqs.~\eqref{conductancevcsimvn}$-$\eqref{conductancevcggvn2}}
\label{appendix_semi}

In this Appendix, we compute the analytic expression of $\mathcal{G}_{\delta p}$ for the two cases $v_c \gtrsim v_n$ (Eq.~\eqref{conductancevcsimvn}) and $v_c \gg v_n$ (Eqs.~\eqref{conductancevcggvn1} and \eqref{conductancevcggvn2}), utilizing a semiclassical approximation such that
the time $t$ in the integrand of Eq.~\eqref{Sup:conductance} is replaced by $ \delta p L/v_c$  except for the time argument $t$ in $G_c(t)$. The approximation is based on the fact that
the dominant contribution to the integrand of Eq.~\eqref{Sup:conductance} comes from a peak structure of $G_c(t)$ near $t=\delta p L/v_c$; the other
peaks from $G_n (t)$ near $t = \delta q L / v_n$ is more monotonous and less important because the scaling dimension ($\delta_n = 1/4$) of the neutral component in the electron tunneling operator is smaller than that ($\delta_c = 3/4$) of the charge component. 
This approximation is applicable when (i) the spatial distance between two interfering neutral components in the QD at $t=\delta p L /v_c$ is much larger than the width $L_{T,n} \propto \hbar v_n / k_B T$ of the neutral components (then $G_n (t)$ is sufficiently monotonous), and (ii) $L \gg 
\hbar v_c / k_B T$ (then $F (t)$, coming from the lead edge, is sufficientlly monotonous).
We focus on the contribution from near $t = \pm L/v_c$ and near $\pm 2L/ v_c$, since
that from larger times $t = \delta p L /v_c$ of $|\delta p| >2$ is much more smaller due to more dephasing.

We first compute $\mathcal{G}_{\delta p = 1}$ in the $v_c \gtrsim v_n$ case with the semiclassical conditions of $k_B T \gg \hbar v_c / L$ and $k_B T \gg \hbar v_n / (L - \Delta L)$, where $\Delta L\equiv L  v_n/v_c$. 
In Eq.~\eqref{Sup:conductance}, the main contribution occurs at $t = \pm L /v_c$. 
Near $t=L/v_c$, we use the following approximations: (1) For the portion of $(\delta p_{\textrm{qm}} , \delta q_{\textrm{qm}}) = (1,-1)$ in the zero-mode part, $\textrm{Re}[G^{0}]  \simeq - \cos(2\pi \Phi/ \Phi_0) \exp (- (L - \Delta L)^2 / (L L_{T,n}) )$; 
cf. Eq.~\eqref{Greenplasmonic_sup}.
(2) For the lead edge part, $F\simeq 4 \pi k_B T L \exp[-2\pi k_B T L / \hbar v_c] / (\hbar v_c ) = (8L / 3L_{T,c}) \exp(-L/L_{T,c}) \exp(-\Delta{L}/L_{T,n})$.
(3) For the QD plasmon part,
\begin{align} \label{plasmonic_sup}
& G_{c}(t) G_{n}(t) \exp\Big(-\frac{3\pi i v_c t }{2L}\Big) \exp\Big(-\frac{i \pi  v_n t }{2L}\Big ) \nonumber \\ \simeq & 2 e^{i \pi / 4} \bigg (\frac{ \sinh (\frac{i\pi k_B T a}{\hbar v_{c}}) }{\sinh \big(\frac{\pi k_B T (L-v_{c}t +ia)}{\hbar v_{c}}\big)} \bigg )^{3/2}\bigg (\frac{a}{L_{T,n}}  \bigg )^{1/2} \nonumber \\ \times  & \exp\big(-\frac{L - \Delta L}{L_{T,n}}\big)
 \exp \big(\frac{(L - \Delta L)^2}{L L_{T,n}}\big). 
\end{align}
We have used $\theta_1 (u,\exp(-\gamma)) \simeq 2 (-1)^{n}\sqrt{\pi /\gamma}\exp [-(u-n\pi)^2 /\gamma]\exp(-\pi^2/4\gamma)\sinh(\pi (u-n\pi)/\gamma)$ for $\gamma \ll 1$.

We compute Eq.~\eqref{Sup:conductance}, merging together (1)-(3), to get Eq.~\eqref{conductancevcsimvn} in the main text,
\begin{align}
 \mathcal{G}_{\delta p =1}&\simeq 
- g_0 \tilde{\gamma} L (k_B T ) ^3
\exp(-\frac{L}{L_{T,c}}) \exp(-\frac{\Delta{L}}{L_{T,n}}) \nonumber
\\ &\times  \exp(-\frac{L - \Delta L}{L_{T,n}})
\nonumber \\ 
&= -g_0 \tilde{\gamma} L (k_B T ) ^3
\exp(-\frac{L}{L_{T,c}})  \exp(-\frac{L}{L_{T,n}}).
\end{align}
Here $g_0 = 16 e^2 \sqrt {2\pi} \pi (a/\hbar v_c)^{13/4} (a/\hbar v_n)^{3/4} (\Gamma(3/4))^2 / (ah)$ is a constant, $\Gamma$ is the Gamma function, and we have used the integral formula of
\begin{equation} \label{integralFM_sup}
\int_{-\infty}^{\infty}dt\bigg (\frac{\sinh(\frac{i\pi a k_B T}{\hbar v_{c}})}{\sinh \big(\frac{\pi k_B T}{\hbar v_{c}}(ia-v_{c}t)\big)} \bigg )^{\frac{3}{2}}  = \frac{2a}{v_c}\Big(\frac{2a k_B T}{\hbar v_c}\Big )^{\frac{1}{2}} \Gamma \big(\frac{3}{4}\big)^2.
\end{equation}
Notice that the factor $\exp (- (L - \Delta L)^2/ (L L_{T,n}) )$ of $\textrm{Re}[G^0]$ (zero-mode part) exactly cancels out $\exp [ (L - \Delta L)^2/ (L L_{T,n}) ]$ from $G_n (t)$ (the plasmonic part). This fact was found in a Luttinger liquid with finite size~\cite{Neutral.Kim}.

We next move to $\mathcal{G}_{\delta p =1}$ in the $v_{c} \gg v_{n}$ case. 
The main contribution to $\mathcal{G}_{\delta p =1}$ also occurs near $t = \pm L /v_c$.
We observe the followings.
(1) Because the portion of $(\delta p_{\textrm{qm}}, \delta q_{\textrm{qm}}) = (1,0)$ in $\textrm{Re}[G^{0}]$ fully vanishes, the dominant contribution comes from the portion of $(\delta p_{\textrm{qm}} , \delta q_{\textrm{qm}}) = (1,-1)$, which leads to
$\textrm{Re} [G^0] \simeq - \cos(2\pi \Phi/ \Phi_0) \exp (- (L - \Delta L )^2 / (L L_{T,n}) )$. (2) $F\simeq 4 \pi k_B T L \exp(-2\pi k_B T L / \hbar v_c) / (\hbar v_c ) = (8L / 3L_{T,c}) \exp(-L/L_{T,c}) \exp(-\Delta{L}/L_{T,n})$. (3) For the QD plasmon part, the same expression is obtained as Eq.~\eqref{plasmonic_sup}, except that $L - \Delta L$ is replaced by $\Delta L$. Merging (1)$-$(3) and using Eq.~\eqref{integralFM_sup}, we obtain Eq.~\eqref{conductancevcggvn1} in the main text,
\begin{align} \label{smallvsFirstHarmonics_sup}
 \mathcal{G}_{\delta p =1}&\simeq 
-g_0 \tilde{\gamma} L (k_B T)^3  \exp(-\frac{L}{L_{T,c}}) \exp(-\frac{\Delta{L}}{L_{T,n}}) \nonumber \\ &\times  
 \exp(-\frac{\Delta L}{L_{T,n}})
 \exp[-\frac{(L -\Delta L)^2-(\Delta L)^2}{L L_{T,n}}]
     \nonumber \\ 
  &= -g_0 \tilde{\gamma} L (k_B T)^3  \exp(-\frac{L}{L_{T,n}})\exp(-\frac{L}{L_{T,c}}).
\end{align}

In the same way, we obtain Eq.~\eqref{conductancevcggvn2},
\begin{align} \label{smallvsSecondHarmonics_sup}
\mathcal{G}_{\delta p =2}& \simeq
2g_0 \tilde{\gamma} L (k_B T)^3 \exp(-\frac{2L}{L_{T,c}}) \exp(-\frac{2\Delta L}{L_{T,n}}) \exp(-\frac{2\Delta L}{L_{T,n}}) \nonumber \\ &= 
 2g_0 \tilde{\gamma} L (k_B T)^3 \exp(-\frac{10L}{3L_{T,c}}).
\end{align}

\section{The quasiparticle fluctuation in the bulk}
\label{Appendix_Quasiflucbulk}

We consider quasiparticle fluctuations in the bulk, differentiating the QD bulk from the edge. In the limit of bulk charging energy smaller than temperature, we show that the zero-mode part $G^0$ has the same expression as Eq.~\eqref{zeromodeaverage2_sup} in the presence of quasiparticle fluctuations in the bulk, hence, that the period halving ($h/2e$ oscillation) takes place.


We argue that the zero-mode part of the QD states is characterized by four numbers ($N_{e}$, $N_{n}$, $N_{qp}$, $N_{qp,n}$), when we additionally consider the bulk degrees of freedom.
$N_e$ and $N_{n}$ are necessary to describe excess electrons in the QD. For an additional excess electron, the number $N_e$ of excess electrons increases by 1. This electron is decomposed into 3 additional charge components ($N_c \to N_c + 3$; the charge of the charge component is $e/3$)
and $\pm 1$ additional neutral component ($N_{n} \to N_{n} \pm 1$), as discussed in the main text. $N_{n}$ counts the number of the neutral components by excess electrons in the QD, as in the main text.
On the other hand, $N_{qp}$ and $N_{qp,n}$ are introduced to describe quasiparticle excitations in the QD bulk. When an additional quasiparticle excites in the bulk, the number $N_{qp}$ of quasiparticles increases by 1. This quasiparticle is decomposed into 1 charge component and $\pm 1$ neutral component  ($N_{qp,n} \to N_{qp,n} \pm 1$).
$N_{qp,n}$ counts the number of the neutral components by bulk quasiparticle excitations.

Taking into account quasiparticle fluctuations in the bulk and electron fluctuations inside the QD, the zero-mode part of the QD Hamiltonian $H_{\textrm{D}}^{0}$ is expressed as 
\begin{align} \label{HamiltonianZeroBulk_sup}
& H_{\textrm{D}}^{0} (N_{e},  N_{qp}, N_{n}, N_{qp,n}) \nonumber \\ = &\, \frac{E_c}{12}\bigg(3N_{e}+1-\frac{2\Phi}{\Phi_0} - N_{qp}\bigg)^2 +
\frac{E_n}{4} (N_{n} - N_{qp,n})^2 \nonumber \\ + & \, E_{\textrm{bc}} \bigg(\frac{2\Phi}{\Phi_0} + N_{qp}\bigg)^2,   
\end{align}  
where $E_{\textrm{bc}}$ is the bulk charging energy.
The first (second) term describes the interaction between the charge (neutral) components on the edge while
the third term describes the interaction between the charge components in the bulk. 
The flux dependence in the bulk-charging energy term describes charge accumulation in the bulk as the magnetic flux increases. 
We ignore the interactions between the neutral components in the bulk
because they are of dipole type hence weaker than the interaction terms of the total charge.
We also assume that it is in the Aharonov-Bohm regime neglecting electrostatic coupling between quasiparticles in the bulk and on the edge.

We consider the case that the relaxation time from the QD edge to the bulk is much longer than the QD dwell time of an electron injected from a lead edge, hence, that the electron enters only into the QD edge.
Then, the zero-mode part $\exp({i\sqrt{3/2}\phi_c^{0}}\pm{i\sqrt{1/2}\phi_n^{0}})$ of the electron operators $\Psi_{\pm}(x)$ is evolved by Eq.~\eqref{HamiltonianZeroBulk_sup} as 
\begin{align}
& e^{{i\sqrt{\frac{3}{2}}\phi_c^{0}(x,t)}\pm{i\sqrt{\frac{1}{2}}\phi_n^{0}(x,t)}} \nonumber \\ 
=& e^{iH_{\textrm{D}}^{0}t}e^{{i\sqrt{\frac{3}{2}}\phi_c^{0}(x)}\pm{i\sqrt{\frac{1}{2}}\phi_n^{0}(x)}}
 e^{-iH_{\textrm{D}}^{0}t}\nonumber \\
=&\exp\Big(i\sqrt{\frac{3}{2}}\phi_c^{0}(x)\pm i\sqrt{\frac{1}{2}}\phi_n^{0}(x) \nonumber \\ +&\frac{i \pi v_c t}{L} (3N_{e}+1-\frac{2\Phi}{\Phi_0} - N_{qp} )\pm \frac{i \pi v_n t}{L}  (N_{n} - N_{qp,n})\Big). 
\end{align}
And, the zero-mode part $G^0(t)$ of the Green's function $\big\langle  \hat{\Psi}_{\pm}^{\dagger}(x_{\mathcal{L}},t) \hat{\Psi}_{\pm}(x_{\mathcal{L}},0)\big\rangle$ is written as 
\begin{widetext}
\begin{align}
 G^0&\propto \Big\langle \exp \Big[\frac{i \pi v_{c}t}{L}  \Big(3N_{e}-\frac{2\Phi}{\Phi_0}+1-N_{qp}\Big)  \Big]
\exp(\pm\frac{ i\pi v_{n}t }{L} (N_{n} - N_{qp,n}) ) 
\Big\rangle \nonumber  \\
& =  \bigg \{
\sum\limits_{\delta p_{\textrm{qm}}, \delta p', \delta q_{\textrm{qm}} = -\infty}^{\infty} e^{-i\pi \delta p' /3} (1+(-1)^{\delta p_{\textrm{qm}}+\delta q_{\textrm{qm}}})(1+(-1)^{\delta p'+\delta q_{\textrm{qm}}})\exp(2\pi i \delta p_{\textrm{qm}} \Phi/ \Phi_0) \nonumber \\& \times
\exp \big [-\frac{3 \pi^2 k_B T }{E_c}\big(\frac{v_{c} t}{L} + \frac{\delta p'}{3} \big)^2 \big] \exp \big[-\frac{\pi^2 k_B T}{E_n}\big( \delta q_{\textrm{qm}} \pm \frac{v_{n} t}{L}  \big)^2 \big] \exp \big[-\frac{\pi^2 k_B T}{4E_{\textrm{bc}}}\big( \delta p_{\textrm{qm}} - \frac{\delta p'}{3} \big)^2 \big] \bigg \} \nonumber \\
& \bigg/ \bigg \{
\sum\limits_{\delta p_{\textrm{qm}}, \delta p', \delta q_{\textrm{qm}} = -\infty}^{\infty}  e^{-i\pi \delta p' /3} (1+(-1)^{\delta p_{\textrm{qm}}+\delta q_{\textrm{qm}}})(1+(-1)^{\delta p'+\delta q_{\textrm{qm}}})\exp(2\pi i \delta p_{\textrm{qm}} \Phi/ \Phi_0) \nonumber \\& \times
\exp \big [-\frac{3 \pi^2 k_B T }{E_c}\big(\frac{\delta p'}{3}\big)^2 \big] \exp \big[-\frac{\pi^2 k_B T\delta q_{\textrm{qm}}^2 }{E_n}\big] \exp \big[-\frac{\pi^2 k_B T}{4E_{\textrm{bc}}}\big( \delta p_{\textrm{qm}} - \frac{\delta p'}{3} \big)^2 \big] \bigg \}.
\label{ZeromodeaverageBulk_sup}  
\end{align}
\end{widetext}
We applied the Poisson summation formula as in Eq.~\eqref{zeromodeaverage2_sup}. 
At temperature much higher than
the bulk-charging energy, the portion of $\delta p' = 3 \delta p_{\textrm{qm}}$ survives, resulting in the same expression as Eq.~\eqref{zeromodeaverage2_sup}.



\section{Confining potential and Coulomb interaction}
\label{appen_Coulomb}

When the edge potential is smooth enough~\cite{Neutral.Wang, Chamon2}, we below show that $v_c$ is much larger than $v_n$ and $v$; see Eq.~\eqref{Hamiltoniandot} in the main text.
In the absense of Coulomb interaction, the edge Hamiltonian at $\nu=2/3$ has the form of
$H_{\text{con}}=\frac{1}{4\pi} \int dx [v_{1,\text{con}} (\partial_x \phi_1)^2 + 3v_{2,\text{con}}(\partial_x \phi_2)^2]$ in terms of
the velocities of the original edge modes, $v_{1,\text{con}}$ and $v_{2,\text{con}}$, which are solely determined by the edge confining potential. 
When the Coulomb interaction is tuned on, 
it leads to an interaction Hamiltonian $H_{\text{int}}=\frac{v_{\text{int}}}{4\pi} \int dx  (\partial_x (\phi_1 + \phi_2))^2 $, which counts the interaction by
the total charge density $\partial_x (\phi_1 + \phi_2)/(2 \pi)$; here, we assume that 
$v_{\text{int}}$ is indepedent of momentum, 
(which is valid when the Coulomb interaction is short ranged due to the screening by gates).
Then, the total Hamiltonian is
$H_{\textrm{D}} = H_{\text{con}}+H_{\text{int}} = \frac{1}{4\pi} \int dx [v_{c} (\partial_x \phi_c)^2 + v_{n}(\partial_x \phi_n)^2+ v\partial_x \phi_c \partial_x \phi_n]$, where $v_c = 3 v_{1,\text{con}} /2+  v_{2,\text{con}} /2+ 2v_{\text{int}}/3$, $v_n =  v_{1,\text{con}}/2+ 3 v_{2,\text{con}}/2$, and $v = -\sqrt{3}(v_{1,\text{con}}+ v_{2,\text{con}})$. This shows 
that  $v$ and $v_n$ are much smaller than $v_c$, if the edge potential is smooth enough such that $v_{\text{int}} \gg v_{1,\text{con}}$, $v_{2,\text{con}}$.

\end{document}